\documentstyle[prd,aps]{revtex}

\def\ap{\approx}
\def\pa{\parallel}

\def\and{\quad\mbox{and}\quad}
\def\dand{\qquad\mbox{and}\qquad}
\def\be{\begin{equation}}
\def\ee{\end{equation}}
\def\ba{\begin{eqnarray}}
\def\ea{\end{eqnarray}}

\title{Curvature Radiation by Ultrarelativistic Protons\footnote{%
         Accepted for publication in {\it Physics Letters B}}}

\author{V. Berezinsky $^{a,b}$, A. Dolgov $^{c,d}$
                 and M. Kachelrie{\ss} $^{a,e}$}

\address{$^a$  INFN, Laboratori Nazionali del Gran Sasso,
               I--67010 Assergi (AQ), Italy }
\address{$^b$ Physikalisches Institut der Universit\"at Bonn,
              D--53115  Bonn, Germany  }
\address{$^c$ Institut for Experimental and Theoretical Physics,
              117259 Moscow, Russia }
\address{$^d$ Dept. Fisica Teorica, Universidad de Valencia,
              E--46100 Burjasot, Spain }
\address{$^e$ Theoretische Physik I, Ruhr-Universit\"at Bochum,
               D--44780  Bochum, Germany  }

\begin{document}

\maketitle
\date{March 3, 1995}

\begin{abstract}
We study pion curvature radiation by a proton, i.e.
pion emission by a proton moving along a curved trajectory.
We suggest an approximate semiclassical solution
and the exact solution for which we assume that a proton moves in a
fictitious magnetic field with the Larmor
radius equal to the curvature radius of the real
trajectory. As possible application we consider the pion radiation by
ultrahigh energy protons moving along curved magnetic field lines.
Such situation can occur in the magnetosphere of a young pulsar,
in the magnetosphere of the accretion disk around a black hole, and in
the vicinity of a superconducting cosmic string.
The decay products of these pions,
such as high energy photons or neutrinos,
can give the observable consequences  of the considered mechanism.
\end{abstract}


\section{Introduction}

According to the commonly used terminology, curvature radiation is
electromagnetic radiation
by a charged particle moving along a curved trajectory. Naturally, the
implied mechanism can be generalized to the emission of any kind of
particles. A simple example is pion curvature radiation by a proton.
Protons can move along a curved trajectory in the gravitational
field, e.g. in the vicinity of a black hole. Another example is the
movement of a proton
along a magnetic field line in the case of a strong field, e.g.
in the magnetosphere of a pulsar or
of an accretion disc. The transverse momentum of a proton in this case is
suppressed due to synchrotron radiation.

This formulation of the problem is semiclassical by its nature:
we define the interaction between particles in a quantum way,
but introduce at the same time the classical concept of a trajectory,
which does not exist in quantum field theory.
We suggest here some solutions to the problem for the case of $\pi^0$
emission by a proton and indicate possible implications.

\section{Kinematics and dynamical parameter of the problem}

To make a quantitative insight into the problem let us consider
a proton moving along a curved trajectory with local curvature
radius $R_C$. We introduce the coherent interaction length, $l_{int}$,
along the trajectory on which the pion is radiated. The four-momentum
of the proton at the begining and the end of $l_{int}$ is denoted
by $(E_i , \vec p_i)$ and $(E_f , \vec p_f)$ respectively,
and that of a pion by
$(\epsilon , \vec k)$. We describe the effect of the curved trajectory
in the interaction region by the longitudinal and transverse
(relative to $\vec p_i$) momenta $q_\pa$ and $q_\bot$ given by some
external field and neglect
the change of proton energy $(|q| \ll |p|)$. Then we have
\begin{equation} \label{e}
 E_f =E_i - \epsilon
 \; , \qquad
 \vec p_{f}= \vec p_{i}- \vec k + \vec q  \; .
\end{equation}
The interaction length is determined  by $q_\pa$,
$l_{int} \ap 1/ q_\pa$. From the set of equations
\be
 q_\bot \ap p_i \theta \; , \qquad
 q_\pa \ap p_i (1-\cos\theta )\ap \frac{1}{2} p_i \theta^2  \and
\theta \ap l_{int}/ R_C \ap 1/(q_\pa R_C)   \; ,
\ee
we find $q_\pa$ and $q_\bot$ as
\be    \label{q}
 q_\bot \ap m \left( \frac{p_i^2}{m^3 R_C} \right)^{1/3}
\quad\mbox{and}\quad
 q_\pa \ap \frac{1}{2}\frac{q_\bot^2}{p_i}  \; ,
\ee
where $m$ is the proton mass.

Naively, the natural parameter of the problem is $1/(mR_C)$.
However, we shall choose as a parameter
the quantity proportional to this value, namely the quantity
in the brackets of Eq. (\ref{q}):
\be     \label{chi}
 \chi  = E_i^2 / (m^3 R_C)  \; .
\ee
The motivation for this choice is given by the connection with
the fictitious magnetic field method. Let us assume that the
considered part of the trajectory is caused by a fictitious magnetic
field $B$ perpendicular to the trajectory plane. This means that the
Larmor radius $R_L= E/(eB)$ equals
the local curvature radius $R_C$. Then Eq. (\ref{chi}) gives
\be    \label{chib}
 \chi = (E_i / m)  (B_\bot /B_{cr})  \; ,
\ee
where $B_{cr}=m^2/e$ and $\chi$ is the standard parameter \cite{ri,so}
\be    \label{chiem}
  \chi =e \sqrt{(F_{\mu\nu}p^\mu)^2}/m^3
\ee
relevant for processes in strong electromagnetic fields.
In this paper we shall confine ourselves to the quantum case $\chi \gg 1$
when the curvature radius is small, $R_C \ll E^2/m^3$.

Two remarks are in order.

The critical magnetic field in our problem is connected with the proton
mass and is equal to $B_{cr} \approx 1.5\cdot 10^{20}\;$ G.

One should distinguish between the parameter  $\chi$ for the curvature
radiation given by Eq. (\ref{chi})
and the parameter $\chi_{em}$ relevant for
the real magnetic field in the considered system. The latter is given by
Eq. (\ref{chiem}) and results in Eq. (\ref{chib}) with
$B_{\perp}$ being the component of the {\em real} magnetic field
$B$ perpendicular to the proton
momentum. The difference between these two parameters,
$\chi$ given by Eq. (\ref{chi}) and $\chi_{em}$ given by
Eq. (\ref{chib}), is very essential for a
proton moving along a curved magnetic field line. The synchrotron
radiation of the proton in the real magnetic field results in the loss of
transverse momentum in the regime when $\chi_{em} \ll 1$:
\be      \label{loss}
\frac{1}{p_{\perp}}\frac{dp_{\perp}}{dt}=
\frac{2}{3}\frac{e^2m}{\gamma}\chi_{em}^2      \; ,
\ee
where $\gamma$ is the Lorentz factor of the proton.
For $R_C < 10^8 \;$cm the lifetime given by (\ref{loss}) in the
regime $\chi_{em}\ll 1$ is shorter than the characteristic time of
flight and therefore a proton moves
along the magnetic field line \cite{su}.
At the same time the curvature radiation
parameter (\ref{chi}) can be larger (or much larger) than 1, providing
thus pion production due to curvature radiation. In other words,
this movement is characterized by $\chi_{em} \ll 1$ and $\chi > 1$
(or $\chi \gg 1$).

\section{Semiclassical solution}

Let us  consider a proton moving along a curved trajectory.
As it is described in the Introduction we single out the interaction
region according to Eq. (\ref{q})
and take the wavefunctions of the
proton at the boundaries as plane waves $\psi_i$ and $\psi_f$.
Then the matrix element describing the interaction is
\be
 M =  g  \bar\psi_f (p_f) \gamma^{5}
 \psi_i (p_i) \phi (k)  \; ,
\ee
where $p_i ,p_f$ and $k$ are determined by Eq. (\ref{e}).
The probability of the transition is calculated in the standard way as
\begin{equation}      \label{}
 d\Gamma =  \frac{(2\pi)^4}{2E_i}\; \frac{1}{2} \; | M |^2  \;
 \frac{ d^3 k}{(2\pi)^3 2\epsilon}\; \frac{ d^3 p_f}{(2\pi)^3 2E_f} \;
 \delta ( E_i - E_f - \epsilon )
 \delta ( \vec p_{i} - \vec p_{f}- \vec k + \vec q )  \; .
\end{equation}
In the evaluation of the spinor traces we use the kinematical
relations given by Eqs. (\ref{e}) and (\ref{q}).
As a result we obtain
\begin{equation}      \label{semi}
  \Gamma =  \frac{3g^2}{64 \pi E_i} \; q_\bot^2
  = \frac{3}{16} \;\alpha_s \;\frac{m^2}{E_i}\:\chi^{2/3} \; ,
\end{equation}
where $\alpha_s=g^2/4\pi$ and $\chi$ is given by Eq. (\ref{chi}) or
in terms of the fictitious magnetic field by Eq. (\ref{chib}).

\section{Fictitious magnetic field method}

In this approach we assume that the curved trajectory is caused by a
fictitious
magnetic field $B$ which is chosen so that $R_L =R_C$. Therefore
we now have to calculate pion radiation of a proton in a strong
magnetic field. This process was already considered in Ref. \cite{zh}
using the crossed field method \cite{ri} (see also Ref. \cite{gi}).
Here, we shall calculate the probability of this process describing it
as a transition  of a proton from the Landau level $N_i$ with polarization
$\tau_i = \pm 1$ into the Landau level $N_f$ due to the interaction
${\cal L}_{int}= g\bar\psi_f \gamma^{5} \psi_i \phi$.
The transition probability
for a proton from the Landau level $N_i$ to $N_f$ is given by
\begin{equation}      \label{exakt}
 \Gamma^{N_i \to N_f}_{\pm} =
 \frac{\alpha_s}{16^2}\sum_{\tau_f}\int_0^{\pi} d\theta \sin\theta \;
  k \; |{\cal M}|^2  \; ,
\end{equation}
where $\theta$ is the angle between the pion momentum $\vec k$ and the
magnetic field.
The matrix element squared $|{\cal M}|^2$ is given in terms
of $I$-functions. For the very large
Landau quantum numbers considered here,
we can use an asymptotic expansion in terms of modified
Bessel functions (for the notations used here and
hereafter see Ref. \cite{so}):
\begin{equation} \label{iasym}
I_{N_f ,N_i}\left(x \right) =
     \frac{\left(1+\xi y\right)^{1/2}}{\pi \sqrt{3}} \; \sqrt{\mu}
      K_{1/3} (z)
\end{equation}
where
\begin{equation}
x=\frac{k_\bot^2}{2eB}
\quad\ ,\quad
z=\frac{2}{3} \left(x_0^2 \, N_i N_f \right)^{1/4}
\left( 1- \frac{x}{x_0} \right)^{3/2}
\quad\ , \quad
x_0 = \left( \sqrt{N_i} - \sqrt{N_f} \right)^2
\end{equation}
\be
 \xi = \frac{3}{2}\frac{B}{B_{cr}}\frac{E_i}{m} = \frac{3}{2}\chi
 \quad , \quad
 y= \frac{2E_i \epsilon \mu^{3/2}}{3eB(1-\epsilon/E_i )}
\ee
\be
 \mu = 1-\frac{2N_i eB}{E_i^2} \sin^2\theta
  \dand
 \mu_0 = 1- \frac{2N_i eB}{E_i^2}  \; .
\ee

Furthermore, since pions are emitted mostly in the forward direction,
i.e. the transition probability
is strongly peaked at $\theta =\pi/2$, we can approximate
all functions except functions of $1-x/x_0$
by their values at $\theta =\pi/2$.
Then we take the ultrarelativistic limit
of the normalization constants of the spinors
and obtain
\ba   \label{M-asy}
 \lefteqn{| {\cal M}|^2 =
    \frac{1+\xi y}{3 \pi^2}  \Biggl[
 \frac{1}{2} \left( 1-\tau_i \tau_f \right) D^2 \mu K_{1/3}^2  (z)}
\nonumber\\ & &
  - \frac{\xi y}{2} \;
  \bigl[ (1-\tau_i \tau_f ) D^2
       - \tau_i \left( 1-\tau_i \tau_f \right) D (E-1)
  \bigr] \mu^{3/2} K_{1/3} (z) K_{2/3} (z)
\\ &  &
  + \frac{(\xi y)^2}{8} \;  \bigl[
  (1-\tau_i \tau_f)(1-2E+D^2+E^2) -
  2\tau_i \left( 1-\tau_i \tau_f \right) D (E-1)
  \bigr] \mu^2 K_{2/3}^2 (z)  \Biggr]  \; ,
\nonumber
\ea
with $D=m/(2E_f)- m/(2E_i)$ and $E=m/(4E_i E_f)$.
Note that because of the pseudoscalar nature of the interaction
only spin-flip transitions occur in the ultrarelativistic
case.

Using now some identities for the Bessel functions
and summing over the polarization states of the proton
we obtain for the spin averaged emission rate
\begin{equation} \label{av}
 \Gamma^{N_i \to N_f} =
 \frac{\alpha_s}{4} \; k  \;
 \frac{y \xi^2 \mu_0^{5/2} (1+\xi y)}{3\sqrt{3}\pi}
 \left( \int_y^{\infty}dx \: K_{5/3}(x) + K_{2/3}(y) \right)  \; .
\end{equation}
For the case $\xi \gg 1$ we can use for the
functions $K_\mu (y)$ their asymptotic expression
$K_\mu (y) = 2^{\mu-1}\Gamma(\mu) y^{-\mu}$ for $y\to 0$. Then,
summing over all final Landau levels yields the
total decay width $\Gamma$
\be     \label{asy}
 \Gamma =
 \frac{2^{2/3}\Gamma^2 (\frac{2}{3})\Gamma(\frac{1}{3})}{18\sqrt{3}\pi}
 \alpha_s \, \frac{m^2}{E_i} \: \xi^{2/3} \ap
 \frac{1}{12.6}\, \alpha_s \,\frac{m^2}{E_i}\: \xi^{2/3}  \; .
\ee
Our width (\ref{asy}) differs only by a numerical factor from the
result of Ref. \cite{zh}. This factor, $9\Gamma (2/3) \Gamma (1/3)/(
2\sqrt{3}\pi)\ap 3.0$, is however rather large.

On the other hand, the agreement between our semiclassical solution
(\ref{semi}) and the Eq. (\ref{asy}) is surprisingly good. Both methods
give not only the same functional dependence on $E_i$ and $\chi$, but
differ only by a numerical factor 1.8. Because of the uncertainities
in the kinematical relations used in the semiclassical calculation
we estimate the uncertainity in the decay width by a factor $4-5$.
However, for two other processes in strong magnetic fields,
$\nu\to\nu +e^++e^-$ and $\gamma\to e^++e^-$, we obtained excellent
agreement between both methods.


\section{Astrophysical applications}

The pion curvature radiation needs protons of extremely high energies.
{}From condition $\chi \geq 1$ one obtains from (7)
\[
E_i \geq6.5\cdot10^9 R_6^{1/2}~ {\rm GeV},
\]
where $R_6=R_C/10^6~{\rm cm}$ is the curvature radius in units of
$10^6~$cm.
In cosmic rays the particles are observed with energies up to
$2\cdot10^{11}~$GeV. The particles with the highest energies are most
probably protons. In the strong magnetic field inside a source protons of
these energies move along the magnetic field lines \cite{su}.
Since the proton momentum is essentially parallel to the magnetic field
the electromagnetic parameter,
$\chi_{em} \ll 1$ and pion production in the strong magnetic field is
absent. However, the curvature parameter (7) can be large if the curvature
radius of the magnetic field line is sufficiently small and therefore the
pion curvature radiation is efficient. Decays of pions result in the
production of gamma and neutrino radiation.

An example can be given by young pulsar.

Independently from the concrete model of pulsar magnetospheres, the maximum
potential $\phi_{max}$ on the surface of of the pulsar is \cite{Be90}:
\be           \label{phi}
\phi_{max} \geq8.3 \cdot 10^{17}(\Omega/10^3~{\rm s}^{-1})^2
(B_s/10^3~{\rm G})(R_s/10^6~{\rm cm})^3~{\rm V},
\ee
where $R_s$ is the star radius, $\Omega$ is the angular velocity and
$B_s$ is the surface magnetic field.

Therefore, pulsars with $\Omega \geq 3 \cdot 10^3~{\rm s}^{-1}$ can, in
principle, accelerate protons to energies higher than $1\cdot
10^{10}~$GeV. If this potential drop occurs nearby the surface of the
pulsar (which is actually not the case of the modern models)
the ultrahigh energy protons escape from the pulsar magnetosphere mostly
along the first-open magnetic field line \cite{Be90}. The
curvature radius of this line is
\be
 R_C = 4/3 \; \left( R_S R_L \right)^{1/2}  \; ,
\ee
where $R_L=c/\Omega$ is the radius of the light cylinder.
For $\Omega = 3\cdot 10^3\;$s$^{-1}$ and $R_S = 1.2\cdot 10^6 \;$cm,
one obtains from Eq. (\ref{phi}) that the energy of accelerated protons
could be higher than $1.3\cdot 10^{10}\;$GeV and the parameter (\ref{chi})
is $\chi \geq 1$.

Another example is an accretion disc, with a strong magnetic field, around
a black hole with a mass $M\sim M_\odot$. The potential at the inner edge
of the disc is \cite{bl}
\be
 \phi = \frac{1}{c} \int \vec E d\vec l =
 \frac{1}{c} \int v(r) B_\phi (r)  dr =
 \frac{1}{\sqrt{6}}\: B_0 r_0 \ln \frac{R}{r_0}  \; ,
\ee
where $r_0$, $R$ and $B_0$ are the radii of the inner and outer edges of
the disc and the magnetic field at the inner edge, respectively.
Taking $E_i\sim e\phi$, $r_0\sim R_C \sim 10^6 \:$cm and $B_0 \sim 10^{10}\:$
G we again obtain $\chi\geq 1$.

Finally, let us discuss superconducting cosmic strings.
Cosmic strings are topological defects caused by symmetry breaking in
the early universe. Under  certain assumptions
the string can become superconducting \cite{Wi85}.
Due to the expansion of the
universe the primordial magnetic field induces an electric current $J$ in
the string which grows up to the maximal value
 $J_{max}\ap eM/2\pi$, where $M$ is the scale of the GUT symmetry
breaking. Starting from this moment, the particles with GUT masses are
evaporating from the string. Decays of these particles produce leptons and
hadrons with maximum energies up to
$M_{GUT}\ap 10^{16}~$GeV.

The string current produces a magnetic field with Biot-Savart
geometry. The initially produced particles have momenta perpendicular
to the magnetic field. Therefore, at a first stage the processes
in the strong magnetic field dominate \cite{Be89}.
After losing energy, particles are captured in circular orbits
\cite{Pe91} and the
pion curvature radiation becomes one of the main energy loss mechanisms of
high energy protons.

\section{Conclusions}

We have studied the curvature radius of pion by a proton. The parameter
of this problem is $\chi = E_i^2 / (m^3 R_C)$,
where $R_C$ is the curvature radius of the proton trajectory and $E_i$
is the proton energy.

For the case $\chi\gg 1$ we found the quasiclassical solution assuming
that in the region of coherent interaction the trajectory gives the
additional
momentum $\vec q$ to the proton. The parallel component of this momentum,
$\vec q_\pa$, determines the length of coherent interaction
$l_{int}\sim 1/q_\pa$. Then the perpendicular momentum, $q_\bot$, is found
as $q_\bot\ap p_i \theta$, where $p_i$ is the proton momentum and
$\theta\ap l_{int}/R_C$. The wavefunctions of the proton at the boundaries
of the interaction region, $l_{int}$, are taken as plane waves and the
matrix element is calculated as $g\bar u (p_f) \gamma_5 u(p_i) \phi$
with $p_f$ and $p_i$ related as $\vec p_f = \vec p_i - \vec k + \vec q$,
where $\vec k$ is the pion momentum. The width of the proton decay,
$p\to p+\pi^0$ is given by
\be
 \Gamma \ap 0.19 \,\alpha_s \, m^2 \, \chi^{2/3} / E_i  \; .
\ee
We estimate the uncertainities of this method by a factor $4-5$.

The other way of calculation is given by the method of fictitious
 magnetic field.
Here we assume that the proton moves along the
trajectory, due to a magnetic field, so that locally
$R_L =R_C$, where $R_L$
is the Larmor radius of the trajectory. We calculated the width of proton
decay in this case as a probabilty of a transition between different
Landau levels and found
\be
 \Gamma \ap 0.104 \,\alpha_s \, m^2 \, \chi^{2/3}/ E_i
\ee
in reasonable agreement with the quasiclassical method.

However, the early calculations of Ref. \cite{zh} for the $p\to p+\pi^0$
decay in the strong magnetic field, made using a different approach,
differ from our calculations by factor 3.

As applications we consider the astrophysical sources where high
energy protons are moving along magnetic field lines. Since the proton
momentum is essentially parallel to the magnetic field, the electromagnetic
parameter $\chi_{em}\ll 1$, while the curvature radiation parameter can be
$\chi > 1$ (in the case of sufficiently small $R_C$). These conditions
are met in such sources as young pulars, accretion dics around black holes,
and superconductive cosmic strings.

\acknowledgements

We are grateful to A. I. Nikishov for comments and for pointing
out Ref. \cite{zh}.
V.B. thanks the Institute of Physics of the Bonn University for
hospitality and the Humboldt Foundation for the award of 1991.
M.K. acknowledges a grant by Deutscher Akademischer Austauschdienst.



\end{document}